\documentclass[pra,twocolumn,superscriptaddress,amsmath,amssymb,aps,floatfix]{revtex4-2}

\usepackage{graphicx}
\usepackage{dcolumn}
\usepackage{placeins}
\usepackage{bm}
\usepackage{amsmath,amssymb}
\usepackage{wrapfig}
\usepackage{setspace}
\usepackage[caption=false]{subfig}
\usepackage{ragged2e}
\DeclareCaptionJustification{justified}{\justifying}
\usepackage{amsfonts}
\usepackage{epsfig}
\usepackage{hyperref}
\usepackage{adjustbox}
\usepackage{rotating}
\usepackage{float}
\usepackage{dblfloatfix}
\usepackage{blindtext}
\usepackage{multirow}
\usepackage{bbold}
\usepackage{import}
\usepackage{braket}
\usepackage{physics}
\usepackage{verbatim}
\usepackage{listings}
\usepackage{xcolor}
\usepackage{xr-hyper}
\usepackage{cleveref}
\usepackage{newfloat}
\usepackage{caption}
\usepackage{subcaption}
\DeclareFloatingEnvironment[name={Fig S},fileext=lof]{suppfigure}
\usepackage[margin=0.6in]{geometry}
\begin{document}
\title{Quantum cryptographic protocols with dual messaging system via 2D alternate quantum walk of a genuine single-photon entangled state}
\author{Dinesh Kumar Panda}
\email{dineshkumar.quantum@gmail.com}
\author{Colin Benjamin}
\email{colin.nano@gmail.com}
\affiliation{School of Physical Sciences, National Institute of Science Education and Research Bhubaneswar, Jatni 752050, India}
\affiliation{Homi Bhabha National Institute, Training School Complex, Anushaktinagar, Mumbai
400094, India}
\begin{abstract}
A single-photon entangled state (or single-particle entangled state (SPES) in general) can offer a more secure way of encoding and processing quantum information than their multi-photon (or multi-particle) counterparts. The SPES generated via a 2D alternate quantum-walk setup from initially separable states can be either 3-way or 2-way entangled. This letter shows that the generated genuine three-way and nonlocal two-way SPES can be used as cryptographic keys to securely encode two distinct messages simultaneously. We detail the message encryption-decryption steps and show the resilience of the 3-way and 2-way SPES-based cryptographic protocols against eavesdropper attacks like intercept-and-resend and man-in-the-middle. We also detail the experimental realization of these protocols using a single photon, with the three degrees of freedom being OAM, path, and polarization. We have proved that the protocols have unconditional security for quantum communication tasks. The ability to simultaneously encode two distinct messages using the generated SPES showcases the versatility and efficiency of the proposed cryptographic protocol. This capability could significantly improve the throughput of quantum communication systems.
\end{abstract}

\maketitle
\newpage
\twocolumngrid
\section{Introduction}
Quantum cryptography~\cite{cryp5} has become a sought-after research field due to the surge in security threats to existing public-key cryptosystems and the unmatched potential of quantum computers over classical counterparts. The former exploits the properties of quantum physics in performing cryptographic tasks for secure quantum communication between two parties~\cite{crypt15,cryp5,cryp7,cryp8,bb84,e91}. The BB84 cryptographic protocol (i.e., Ref.~\cite{bb84}) was the first to realize quantum superposition in a cryptographic protocol, wherein the qubits are transmitted in one direction, and classical information exchange is needed to establish the communication between two parties (sender-receiver, or, conventionally Alice-Bob). An eavesdropper (Eve) can be detected from an erroneous key at the receiver's end, and the sender-receiver duo repeats the entire procedure if needed. Ref.~\cite{e91} is the famous E91 cryptographic protocol that uses Bell states to achieve secure communication between the two parties. In this case, Alice and Bob test Bell's inequalities to detect the existence of an eavesdropper. In Ref.~\cite{2way}, a direct communication protocol between Alice and Bob without the use of any entanglement to send messages \{0, 1\}, is discussed exploiting nonorthogonal states in a two-way quantum channel.

Discrete-time quantum walks (DTQWs or QWs) are a potent resource for quantum computing and executing quantum-information-processing tasks~\cite{qwgenqmeasure23,qwunivqi23,qwqudit19,qw93}. Refs.~\cite{crypt15,apanda,me-cb2,cryp13} introduced quantum walks in quantum cryptographic systems. The use of DTQWs in quantum cryptographic protocols~\cite{bb84,e91} has advantages, such as a more straightforward way to generate public-key via a DTQW setup instead of using random number generators and numerous photons with classical communication~\cite{bb84} or entangled spin-$\frac{1}{2}$ particles with random measurements~\cite{e91}. Moreover, the message decryption via DTQW setup is less resource intensive than protocols discussed in~\cite{bb84,e91} as these deploy random measurements at the receiver's end and classical
communication between the receiver-sender duo to decrypt the encoded message. Further, the BB84 protocol~\cite{bb84} lacks entanglement, making the process of detecting Eve's interference laborious, while the protocol in Ref.~\cite{2way} can enable an encode-decode of a single message (either 0 or 1), in one shot. Moving now on to DTQW-based cryptographic schemes, Ref.~\cite{cryp13} showed how a two-particle cyclic-QW could generate random key sequences due to its inherent nonlinear chaotic dynamic behavior, which can be used in image encryption. On the other hand, Ref.~\cite{crypt15} proposed that a cryptographic public-key can be formed from a quantum state generated through cyclic-QW (CQW) dynamics. Few recent studies also use quantum states generated via ordered or periodic 1D CQWs in quantum cryptography; see~Refs.~\cite{me-cb2,apanda}. Ref.~\cite{apanda} uses periodic CQW-states to generate a public key, whereas Ref.~\cite{me-cb2} discussed a secure CQW-based cryptographic scheme via the use of a recurrent maximally entangled state in forming the public key. The generated QW states are cryptographic keys to encrypt and decrypt a single message between the sender-receiver duo. The chosen position Hilbert space of the CQWs in Refs.~\cite{me-cb2,crypt15,apanda} is one-dimensional. Thus, it poses a limitation for encrypting an extended set of messages and enabling the simultaneous encryption of more than one message. It motivates us to design a QW-based cryptographic scheme with more spatial degrees of freedom (DoF), like 2D alternate DTQWs (hereafter referred to as alternate quantum walk or AQW)~\cite{p3,BuschPRL}. This letter shows how to perform cryptographic tasks with two distinct sets of messages that can be securely communicated between a sender-receiver duo. In one shot, two messages can be sent. Furthermore, in our cryptographic scheme, the sender (Alice) need not know the form of the public key (or private key) to perform the dual-message encoding, while the receiver (Bob) can deterministically determine the messages received, i.e., a classical channel is not demanded in our protocol. This, in turn, makes our protocol more efficient and secure for dual messaging.
\section{Motivation}
Our proposal is unique and resource-saving as it uses only a single photon (i.e., a quantum walker/particle), exploiting entanglement between either of the three degrees of freedom (genuine 3-way entanglement) or the two nonlocal degrees of freedom (nonlocal 2-way entanglement) of the photon to enable a dual message encryption-decryption scheme. {The motivation behind our study is that, in one shot, two messages can be sent simultaneously via the same AQW setup, as compared to all other protocols, which are sequential and in there simultaneous encryption of two messages is impossible. Our protocol reduces the resources required to perform the communication as we, for the first time, show that by exploiting two nonlocal degrees of freedom of the photon (or walker), one can encode the two messages.} {Further, we can encode-decode an extended set of messages and enable the simultaneous encryption of more than one message. This is due to the high-dimensional Hilbert spaces for each of the positions: $x$ and $y$ degrees of freedom of the walker (photon). We have bettered the previously used QW setups, which use 1D cyclic graphs and are limited to a low-dimensional Hilbert space and can not perform communication of more than one message (and that even from a small set of numbers).} {The generation of the entangled single photon and the secure encryption-decryption of two messages is carried out via a 2D AQW setup employing a resource-saving single-qubit coin operator (e.g., a Hadamard gate) and an initial separable state. This is another main motivation and important conclusion of our study, that our protocol uses only a single photon to achieve secure communication between the sender-receiver duo over the multi-photon-based existing protocols.} Three-way, as well as nonlocal two-way single particle entangled states (SPES), can be generated via a 2D AQW efficiently~\cite{p3}. The amount of three-way or two-way single-particle-entanglement (SPE) generated can be controlled by tuning the initial state and evolution operator parameters of the AQW dynamics~\cite{p3}. This work proposes a secure communication protocol between the sender-receiver duo using the generated genuine three-way and nonlocal two-way SPES as cryptographic keys. This protocol can encode dual (or two simultaneous) messages using the quantum walker e.g., a single photon. We also provide the necessary proof and show via examples how the proposed cryptographic scheme has foolproof security against eavesdroppers. We then compare it with other quantum-walk-based cryptographic protocols. We also investigate the cryptographic scheme's ability to withstand eavesdropping (like man-in-the-middle, intercept-and-resend attacks).

Below, in Sec.~III, we first detail the AQW dynamics to generate genuine 3-way and nonlocal 2-way SPES. Then, in Sec.~IV we delve into the specifics of our cryptographic protocol using the generated 3-way SPES along with necessary proofs and its resilience (security) against attacks from eavesdroppers, with example messages. Further, a similar cryptographic protocol using the generated nonlocal 2-way SPES is discussed, and a comparison with existing proposals is also made. We then show our scheme's single photon-based experimental realization via a circuit, in Sec.~V. Finally in Sec.~VI, the letter concludes with a summary and outlook.

\section{Generating either three-way or nonlocal two-way SPES via 2D AQW}
A 2D AQW is defined on a tensor product space $H_x\otimes H_y\otimes H_c$ of the position and coin Hilbert spaces. $H_y$ and $H_x$ are the $y$-position and $x$-position Hilbert spaces (infinite dimensional), and $H_c$ refers to the coin Hilbert space (two dimensional) with $\{\ket{0_c},\ket{1_c}\}$ being its computational basis. A photon is a practical example of the quantum walker, wherein its polarization states (such as horizontal $\ket{H}$ and vertical $\ket{V}$ polarization) are mapped into $H_c$ space, and the photon's path DoF into  $H_y$ position space and finally, its OAM (orbital angular momentum) DoF corresponds to the $H_x$ space~\cite{Chandra2022}. The quantum walker, when initially localized at the position $\ket{0_x,0_y}$, with an arbitrary superposed coin state, has the general initial state~\cite{p3}, $\ket{\psi_0}=\ket{\psi(t=0)}$, i.e.,
\begin{equation}
\ket{\psi_0} = \ket{0_x,0_y}\otimes[\cos(\frac{\theta}{2})\ket{0_c} + e^{i\phi}\sin(\frac{\theta}{2})\ket{1_c}].
\label{eq1}
\end{equation}

Here phase $\phi \in [0, 2\pi)$, $\theta \in [0, \pi]$, and the particle (2D alternate quantum walker) evolves via the coin operation, $\hat{C}$, and the two conditional shift operators $\hat{S_x}$ and $\hat{S_y}$ which enable the particle transport in $x$-direction and $y$-direction alternately. The following is a general single-qubit coin operator,
\begin{equation}
\hat{C}(\alpha,\beta,\gamma) =
\begin{pmatrix}
\cos{(\alpha)} & e^{i\beta} \sin{(\alpha)}\\
\sin{(\alpha)}e^{i\gamma} & -e^{i(\beta+\gamma)} \cos{(\alpha)}
\end{pmatrix},
\label{eq2}
\end{equation}
with, $\alpha,\beta,\gamma\in[0,2\pi]$, while 
 the conditional shift operators are given as,
\begin{equation}
\begin{aligned}
\hat{S_x} = & \sum_{j=-\infty}^{\infty}\{\ket{(j-1)_x}\bra{j_x}\otimes\mathbb{1}_y\otimes\ket{0_c}\bra{0_c}\\
&+\ket{(j+1)_x}\bra{j_x}\otimes\mathbb{1}_y\otimes\ket{1_c}\bra{1_c}\},
\\ \text{and }
\hat{S_y} = & \sum_{j=-\infty}^{\infty}\{\mathbb{1}_x\otimes\ket{(j-1)_y}\bra{j_y}\otimes\ket{0_c}\bra{0_c}\\
&+\mathbb{1}_x\otimes\ket{(j+1)_y}\bra{j_y}\otimes\ket{1_c}\bra{1_c}\}.
\end{aligned}
\label{eq3}
\end{equation}The total evolution operator at each time-step ($t$) for the 2D alternate quantum walker is then given by~\cite{p3},

\begin{equation}
\begin{split}
U= (\hat{S}_y.[\mathbb{1}_{xy}\otimes \hat{C}]).(\hat{S}_x
.[\mathbb{1}_{xy}\otimes \hat{C}])=C_yC_x \;,
\label{eq4}
\end{split}
\end{equation}

where, partial evolution operators, $C_y= (\hat{S}_y.[\mathbb{1}_{xy}\otimes \hat{C}])$ and $ C_x=(\hat{S}_x
.[\mathbb{1}_{xy}\otimes \hat{C}])$. $\hat{C}=\hat{C}(\alpha,\beta,\gamma)$ is the coin operator. Both $\mathbb{1}_x$ and  $\mathbb{1}_y$ are $
(2N+1)\times(2N+1)$ identity matrices, while $\mathbb{1}_{xy}$ is a $
(2N+1)^2\times(2N+1)^2$ identity matrix, which are basically the identity operators in $H_x$, $H_y$ and combined $xy$-position Hilbert-space respectively. Thus, the particle (walker) at $t=N$ has a quantum state,
\begin{equation}
\ket{\psi_{t}} = \sum_{i,j=-N}^{N}[h_{i,j}^{(N)}\ket{i_x,j_y,0_c}+v_{i,j}^{(N)}\ket{i_x,j_y,1_c}],
\label{eq5}
\end{equation}
wherein the complex amplitudes $h_{i,j}^{(N)}$ and $v_{i,j}^{(N)}$ are functions of the initial state parameters: $\phi,\theta$ and coin parameters: $\alpha,\beta,\gamma$. The intertwined state amplitudes indicate that these states can be genuine three-way entangled between the coin, $x$-position and $y$-position DoF, and nonlocal two-way entangled between $x$-position DoF and $y$-position DoF of the walker~\cite{p3}.

\begin{figure}[h!]
\includegraphics[width = 9.1cm,height=5.5cm]{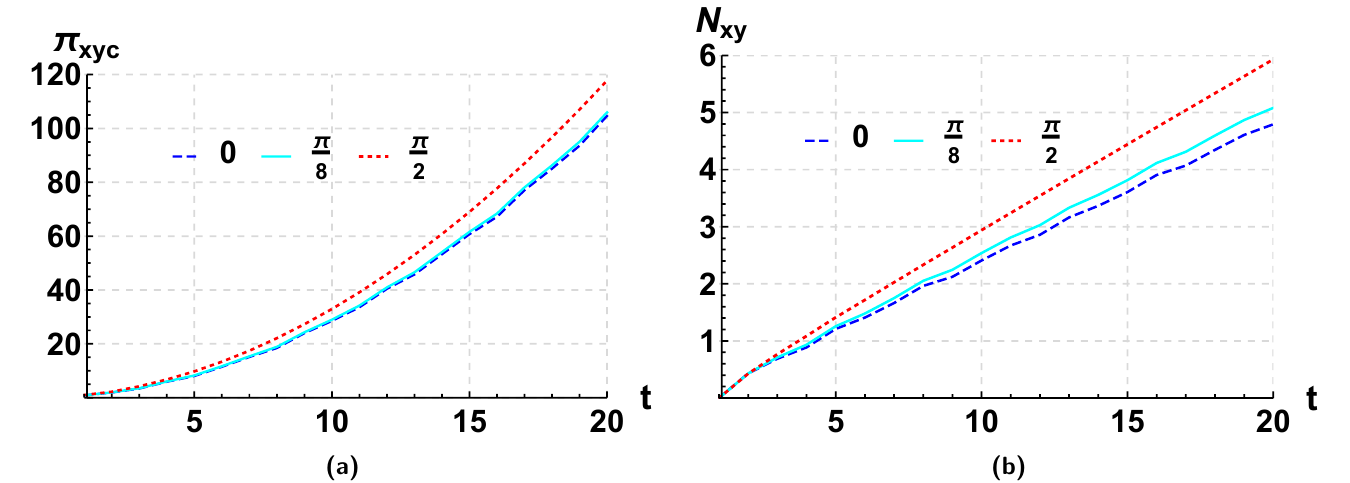}
\caption{(a) $\pi$-tangle $\pi_{xyc}$ vs time steps ($t$) for evolution sequence $M1_xM1_y...$ and with an initially separable state (Eq.~(\ref{eq1})) with $\phi=\pi$, and $\theta=\frac{\pi}{2}$(best result, dotted red),  $\theta=0$(dashed blue), $\theta=\frac{\pi}{8}$(solid cyan). (b) Nonlocal 2-way entanglement negativity $N_{xy}$ vs $t$ for evolution sequence $G1_xG1_y...$ and with the initially separable state with $\phi=\pi$, and $\theta=\frac{\pi}{2}$(best result, dotted red),  $\theta=0$(dashed blue), $\theta=\frac{\pi}{8}$(solid cyan).}
\label{f2}
\end{figure}

In a recent work~\cite{p3}, we performed several numerical simulations to reveal a wide range of spatial evolutions (see, Eq.~(\ref{eq4})) comprising a single arbitrary coin (see, Eq.~(\ref{eq2})), which generate maximal three-way and nonlocal two-way SPE for the separable initial state~Eq.~(\ref{eq1}). We call such evolution sequences optimal entanglers. In addition, the amount of 3-way and nonlocal 2-way SPE so generated can also be controlled by fine-tuning the initial state or coin parameters.

One such optimal entangler for 3-way entanglement (see, Eq.~\ref{eq4}) among numerous other optimal entanglers of Ref.~\cite{p3} is, $M1_xM1_y,$ with $\phi=0,\pi,2\pi$, here the coin $\hat{M1}$ is $\hat{C}(\alpha=\frac{5\pi}{16},\beta=\frac{\pi}{2},\gamma=\frac{\pi}{2})$. Fig.~\ref{f2}(a) shows the generated 3-way entanglement quantified by $\pi$-tangle~\cite{fan,pitang}, $\pi_{xyc}$ (see Appendix~\ref{appa}) via the evolution sequence, $M1_xM1_y...$ with $\phi=\pi$ and $\theta=0,\frac{\pi}{8},\frac{\pi}{2}$. Note that $\pi_{xyc}=0$ in the absence of 3-way entanglement, and the larger the $\pi_{xyc}$ value, the larger the genuine 3-way entanglement in the SPES. From Fig.~\ref{f2}(a), we see that $M1_xM1_y...$ gives largest 3-way entanglement for $\theta=\frac{\pi}{2}$, such as $\pi_{xyc}=2.2070 \;(\text{at } t=2),32.9376\;(\text{at }t=10), 117.7828\;(\text{at }t=20)$.

An optimal entangler  (Eq.~\ref{eq4}) for the nonlocal 2-way entanglement (quantified by the entanglement negativity $N_{xy}$, see Appendix~\ref{appa} and~\cite{neg2002}) from Ref.~\cite{p3} is $G1=\hat{C}(\alpha=\frac{19\pi}{16},\beta=\frac{\pi}{2},\gamma=\frac{\pi}{2})$ with $\phi=0,\pi$. Fig.~\ref{f2}(b) shows the negativity ($N_{xy}$) generated via the evolution $G1_xG1_y$ with $\phi=\pi$ and $\theta=0,\frac{\pi}{8},\frac{\pi}{2}$. Clearly, the evolution $G1_xG1_y...$ yields maximal nonlocal 2-way entanglement for $\theta=\frac{\pi}{2}$ and specifically, at time steps $t=2,10,20$; $G1_xG1_y$ yields correspondingly $N_{xy}=0.4273,2.9398,5.9312$. 

Below, we discuss how this generated 3-way and nonlocal 2-way SPES are utilized in a cryptographic protocol to design quantum-secure public keys and how one can perform the encryption-decryption of two distinct messages simultaneously via $x$-DoF and $y$-DoF of the quantum-particle, between two parties: Alice and Bob.

\section{3-way SPES-based quantum cryptographic protocol}
The 3-way SPE-generating 2D AQW scheme can be exploited to design a cryptography protocol to encode two messages ($m,n$) in the two nonlocal DoFs ($x,y$) of the particle evolving via the AQW. We discuss the protocol considering the spatial evolution sequence $M1_xM1y...$ in the following steps--

\underline{Step 1} (Entangled Public Key Generation): Say Alice plans to send two distinct messages, $m$ and $n$, to Bob securely, with $m,n\in\{-N,-(N-1),...,-2,-1,0,1,2,..., N-1, N\}$. {These are messages each of maximum $\text{log}_2(2 N+1)$ bits, and $N$ is the maximum value for position of the walker in the  $x$ or $y$ direction at time step $t$ of the AQW evolution.} Bob generates the public key as $\ket{\phi_{pk}}=U^t\ket{l_x}\ket{k_y}\ket{q_c}$, where $U=C_yC_x=\hat{S}_y.[\mathbb{1}_{xy}\otimes \hat{M1}].\hat{S}_x
.[\mathbb{1}_{xy}\otimes \hat{M1}]$, see Eq.~(\ref{eq4}). $\ket{l_x}$ and $\ket{k_y}$ are respectively the $x$ and $y$ position states and $\ket{q_c}=\cos(\frac{\theta}{2})\ket{0_c}+e^{i\phi}\sin(\frac{\theta}{2})\ket{1_c}$ is the coin state, of the walker (particle). This public-key is entangled in three DoF i.e., $x$, $y$ and coin DoF, e.g., with $\pi$-tangle $\pi_{xyc}=2.2070$ at $t=2$, $\pi_{xyc}=32.9376$ at $t=10$ and so on, for $\phi=\pi$ and $\theta=\frac{\pi}{2}$ (see Fig.~\ref{f2}(a)). Note: These are maximal values of entanglement obtainable for arbitrary initial state (Eq.~\ref{eq1}) via the AQW dynamics. Bob sends this secure 3-way entangled public key to Alice at any arbitrary time-step ($t$) via a quantum channel.

\underline{Step 2} (Message Encryption): Now Alice encodes the messages $m$ and $n$ respectively in the $x$-DoF and $y$-DoF, via $\ket{\psi(m,n)}=\hat{T}\ket{\psi_{pk}}$, where $\hat{T}=\sum_{i,j=-N}^{N}\ket{(i+m)_x}\bra{i_x}\otimes\ket{(j+n)_y}\bra{j_y}\otimes\mathbb{1}_c$. Then Alice sends the encoded messages via quantum state $\ket{\psi(m,n)}$ to Bob.

\underline{Step 3} (Message Decryption): Bob decodes the encrypted message via the operation $\hat{M}=(U^{-1})^{t}$ on $\ket{\psi(m,n)}$. Here, $U^{-1}=(C_yC_x)^{-1}=C_x^{-1}C_y^{-1}=(\hat{S}_x
.[\mathbb{1}_{xy}\otimes \hat{M1}])^{-1}.(\hat{S}_y.[\mathbb{1}_{xy}\otimes \hat{M1}])^{-1}$, also see Eq.~(\ref{eq4}). It is essential to have $[U,\hat{T}]=0$, i.e., the operators $U$ and $\hat{T}$ commute, for proceeding further with this cryptographic protocol. We provide the detailed proof for the commutation relation $[U,\hat{T}]=0$ in Appendix~\ref{appb}. Thus, Bob obtains, 

\begin{equation}
\begin{aligned}
    \hat{M}\ket{\psi(m,n)}= &  (U^{-1})^t\hat{T}\ket{\psi_{pk}} =(U^{-1})^t\hat{T}U^t\ket{l_x,k_y,q_c}, \\ & = U^{-t}\hat{T}U^1U^{t-1}\ket{l_x,k_y,q_c},\\ &= U^{-t}U^1\hat{T}U^{t-1}\ket{l_x,k_y,q_c},\\ & =...=U^{-t}U^{t-1}\hat{T}U^1\ket{l_x,k_y,q_c},\\ & =U^{-t}U^t\hat{T}\ket{l_x,k_y,q_c},\\ & =\ket{(l+m)_x}\ket{(k+n)_y}\ket{q_c}.
\end{aligned}
\label{eq6}
\end{equation}

The second and third lines of Eq.~(\ref{eq6}) justify the necessity of the commutation relation $[U,\hat{T}]=0$ (proved in Appendix~\ref{appb}) in this protocol. Finally, from the outcome of Eq.~(\ref{eq6}), Bob securely reads the two messages ($m,n$) encoded by Alice. 
The security of the cryptographic scheme against any eavesdropper attack is discussed below.

{\bf{Security of the protocol with example messages:}}
Here, we discuss the abovementioned cryptography scheme and its resilience against eavesdropping attacks, with example messages: $m=1$ and $n=2$.

Let $\ket{l_x}=\ket{0_x}\;, \ket{k_y}=\ket{0_y}\;,\ket{q_c}=\frac{1}{\sqrt{2}}(\ket{0_c} -\ket{1_c})$ which corresponds to the initial separable state shown in Eq.~(\ref{eq1}) with $\phi=\pi$ and $\theta=\frac{\pi}{2}$, and we consider the 2D AQW state at time-step $t=2$, say. 

\underline{Step 1}: The public key generated by Bob is, $\ket{\phi_{pk}}=U^2\ket{0_x}\ket{0_y}\frac{\ket{0_c} -\ket{1_c}}{\sqrt{2}}$, where $U=\hat{S}_y.[\mathbb{1}_{xy}\otimes \hat{M1}].\hat{S}_x
.[\mathbb{1}_{xy}\otimes \hat{M1}]$. 

Thus, $\ket{\phi_{pk}}=U^2\ket{0_x,0_y}\otimes\frac{1}{\sqrt{2}}(\ket{0_c}-\ket{1_c})=\frac{1}{8192}[256u^2(-e^{\frac{i\pi}{2}}v+4\sqrt{2}u^2)\ket{-2_x,-2_y,0_c}+e^{\frac{i\pi}{2}}\{-8e^{\frac{i\pi}{2}}v^2(-4\sqrt{2}+e^{\frac{i\pi}{2}}vu^{-2})\ket{-2_x,0_y,0_c}+(64\sqrt{2}e^{\frac{i\pi}{2}}v^2-8e^{i\pi}v(-32+v^2)u^{-4})u^2)\ket{0_x,-2_y,0_c}-(64\sqrt{2}e^{\frac{3i\pi}{2}}v^2-8e^{2i\pi}v^3u^{-2}+256e^{2i\pi}vu^{-2})\ket{0_x,0_y,0_c}+(\sqrt{2}e^{\frac{3i\pi}{2}}v^4u^{-4}+8e^{2i\pi}v^3u^{-2})\ket{2_x,-2_y,0_c}+(32\sqrt{2}e^{\frac{5i\pi}{2}}v^2+256e^{3i\pi}vu^2)\ket{2_x,0_y,0_c}+(-32\sqrt{2}e^{\frac{i\pi}{2}}v^2+256vu^2)\ket{-2_x,0_y,1_c}
+(8e^{i\pi}v^3u^{-2}-\sqrt{2}e^{\frac{3i\pi}{2}}v^4u^{-4})\ket{-2_x,2_y,1_c}
+(64\sqrt{2}e^{\frac{3i\pi}{2}}v^2+8e^{i\pi}v^3u^{-2}-256e^{i\pi}vu^2)\ket{0_x,0_y,1_c}-(64\sqrt{2}e^{\frac{5i\pi}{2}}v^2+8e^{2i\pi}v^3u^{-2}-256e^{2i\pi}vu^2)\ket{0_x,2_y,1_c}-(32\sqrt{2}e^{\frac{5i\pi}{2}}v^2+8e^{2i\pi}v^3u^{-2})\ket{2_x,0_y,1_c}-(1024\sqrt{2}e^{\frac{7i\pi}{2}}u^4+256e^{3i\pi}vu^2)\ket{2_x,2_y,1_c}\}]$, where $u=\sin(\frac{3\pi}{16}),\; v=\csc(\frac{\pi}{8})$. This public-key quantum state with intertwined complex amplitudes is entangled in the three available DoF: $x$, $y$ and coin, with degree of entanglement $\pi_{xyc}=2.2070$. Bob sends this 3-way entangled public key to Alice via a quantum channel. An eavesdropper (Eve) can interrupt this communication at the stage (step-2) when Alice transfers the encrypted
message state $\ket{\psi(m=1,n=2)}$ to Bob.

\underline{Step 2}: Alice encodes the messages $m=1$ and $n=2$ respectively in the $x$-DoF and $y$-DoF, via $\ket{\psi(m=1,n=2)}=\hat{T}\ket{\psi_{pk}}$, where $\hat{T}=\sum_{i,j=-2}^{2}\ket{(i+1)_x}\bra{i_x}\otimes\ket{(j+2)_y}\bra{j_y}\otimes\mathbb{1}_c$ is kind of a shift operator (Eq.~\ref{eq3}) that causes shift in $x,y$-directions independent of coin state. Then Alice sends this message-containing quantum state $\ket{\psi(m=1,n=2)}$ to Bob.

\underline{Step 3}:  In the absence of Eve, Bob receives $\ket{\psi(m=1,n=2)}$ untampered and decodes the messages by operating $\hat{M}=U^{-2}$ on $\ket{\psi(m=1,n=2)}$. Bob now obtains,  $\hat{M}\ket{\psi(m=1,n=2)}=U^{-2}\hat{T}\ket{\psi_{pk}}=U^{-2}\hat{T}U^2\ket{0_x,0_y}\otimes\frac{1}{\sqrt{2}}(\ket{0_c} -\ket{1_c})=U^{-2}U^2\hat{T}\ket{0_x,0_y}\otimes\frac{1}{\sqrt{2}}(\ket{0_c} -\ket{1_c})=\hat{T}\ket{0_x,0_y}\otimes\frac{1}{\sqrt{2}}(\ket{0_c} -\ket{1_c})=\ket{(0+1)_x,(0+2)_y}\otimes\frac{1}{\sqrt{2}}(\ket{0_c} -\ket{1_c})=\ket{1_x}\ket{2_y}\otimes\frac{1}{\sqrt{2}}(\ket{0_c} -\ket{1_c})$. Finally, from this output, Bob securely reads the two messages ($m=1,n=2$) encoded by Alice: the $x$ and $y$ position kets.

Now, let us consider that Eve is present at Step 2 and can access quantum computers (say, Q-Eve) or classical computers (say, C-Eve). {Eve, however, does not know the private key, i.e., values of $\{U, l_x,k_y,t\}$, thus C-Eve knowing the public key state $\ket{\psi_{pk}}$ and hence the message-containing quantum state $\ket{\psi(m=1,n=2)}$ is well-nigh impossible~\cite{crypt15}. A detailed analytical proof of this fact, incorporating Holevo's theorem and calculations of the Shannon entropy of the private key, along with the mutual information between the private key and Eve's deduction, is provided in Appendix~\ref{appc}. There we show the Shannon entropy of the private key has at least a polynomial overhead over the
von Neumann entropy of the public key as seen by Eve, thereby implying the public key to be secure. Further, the genuine 3-way entanglement (maximal) in public key $\ket{\psi_{pk}}$ (in comparison to a product-state or superposed public key of Refs.~\cite{apanda,crypt15}) makes it even harder for both Q-Eve or C-Eve to extract any information from the public key via POVM measurements~\cite{crypt15}. It is straightforward to see that without the knowledge of the public key (or the private key), the message-containing quantum state $\ket{\psi(m,n)}$ as perceived by Eve is also a completely mixed state, and it is at least as secure as the public key (proven in Appendix~\ref{appc}). Furthermore, as a direct consequence of the fact that a specific private key (which is proven to be secure in Appendix~\ref{appc}) can only enable the decryption of the messages $(m,n)$ from the dual-message encrypted quantum state $\ket{\psi(m,n)}$, Eve figuring out the messages is impossible. This fact mitigates the intercept-and-resend eavesdropping strategy~\cite{qkd-attack2020} wherein Eve tries to measure the signal sent by Alice, i.e., $\ket{\psi(m,n)}$.} Similarly, Q-Eve can not harm, and also her measurements perturb and alter the quantum state $\ket{\psi(m,n)}$ and make the state unusable. This attempt of interception will be recognized immediately by the sender-receiver duo. 
 In addition, the fact that Alice and Bob share the private key $\{U, l_x,k_y,t\}$ takes care of the risk of the receiver's authentication (i.e., a man-in-the-middle attack~\cite{qkd-attack2020}), i.e., whether the receiver is the friend (Bob) or a foe (Q-Eve or C-Eve). Finally, Q- or C-Eve needs to have information on the operator to be applied on $\ket{\psi(m=1,n=2)}$ to retrieve Alice's message(s). Additionally, any Eve figuring out exact $\hat{M}$ (dependent on the secure private key) and retracing back to the state $\ket{\psi(m=1,n=2)}$ is practically impossible, as there exists an infinite number of possible coin operators and their combinations to form an evolution operator via the complex 2D AQW dynamics, also see Appendix~\ref{appc}. Therefore, the proposed quantum cryptographic protocol is secure and resilient against any Eve, whether Eve has access to quantum or classical computers. Although this quantum cryptographic protocol is foolproof, implementing hardware adaptation to isolate quantum systems and employing privacy-amplification is highly recommended~\cite{qkd-attack2020}. These measures can further mitigate deviations from the theoretical prediction in the practical implementation of the quantum cryptographic scheme. 

\begin{widetext}

\begin{figure}[t]
\includegraphics[width=7in,height=4in]{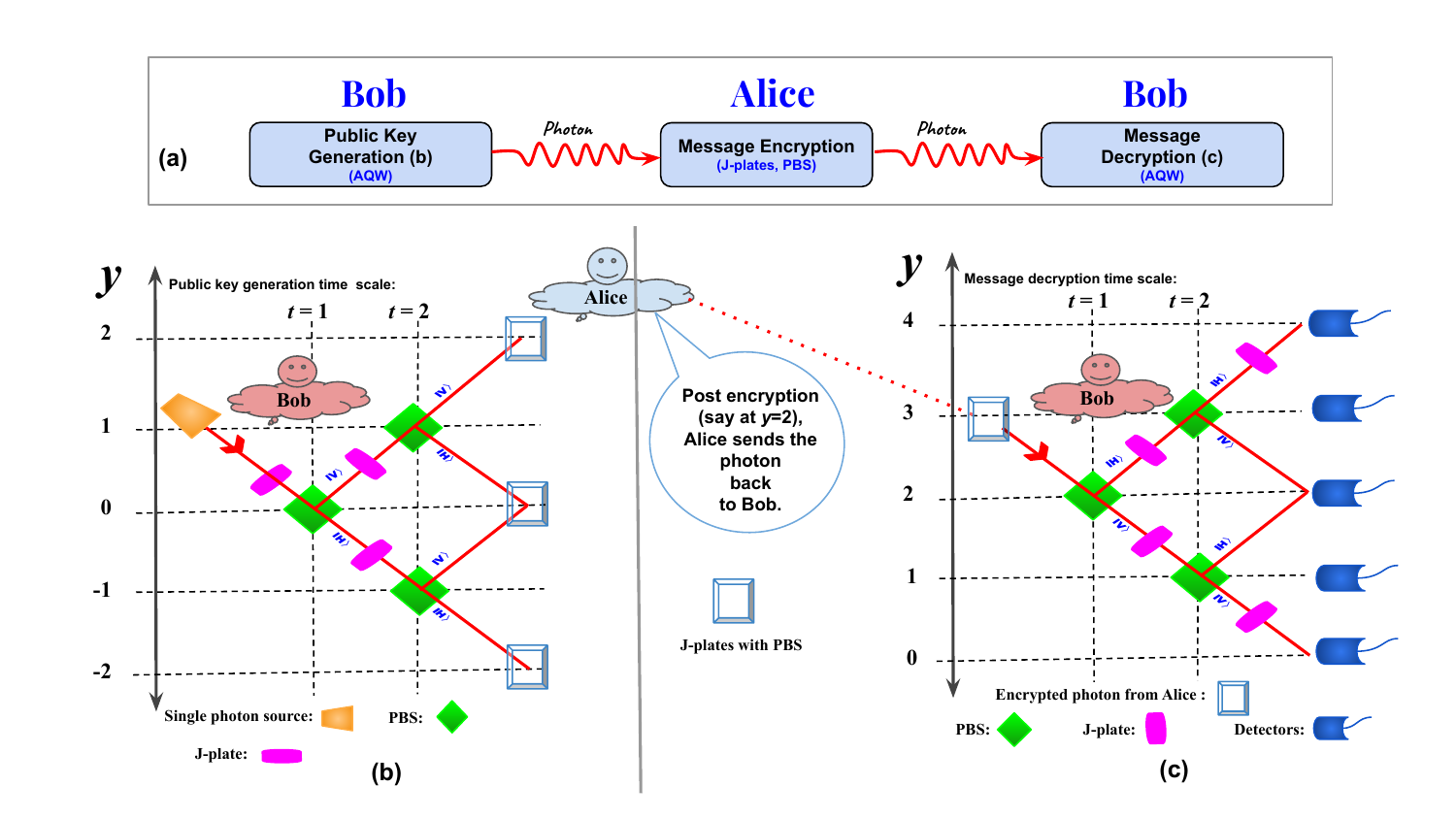}
\caption{Photon-based realization of the quantum cryptography protocol using genuine 3-way single-particle entangled states: (a) the cryptography protocol for secure communication between Bob--Alice--Bob using a photon; (b) Circuit for generation of the public key via AQW using a single photon by Bob and message encryption by Alice; (c) Message Decryption circuit: Bob decrypts the messages from the photon using this circuit. Note that the incoming photon from Alice (blue box) is actually in a superposition of path DoF. Thus, the decryption apparatus (c) is at every probable path of the incoming photon. For instance, at timestep $t=2$ 
with say message $n=0$ (encrypted in the path DoF), the copies of the decryption apparatus are to be at 3 spatial positions, namely $y=-2$, $y=0$ and $y=2$.}
\label{f3}
\end{figure}
\end{widetext}

A quantum cryptographic protocol similar to the above-discussed 3-way SPES-based cryptographic protocol can be designed, utilizing the nonlocal 2-way SPES generated via the 2D AQW dynamics (Fig.~\ref{f2}(b)) as public keys. See Appendix~\ref{appd} for message encryption-decryption steps of the protocol and Appendix~\ref{appe} for its security analysis against eavesdroppers.
Although the cryptographic protocols based on 3-way SPES and nonlocal 2-way SPES have quite the step 2 (message encryption), step 1 (entangled public key generation) and Step 3 (Message Decryption)  of the two protocols are different. In the latter, step 1 involves a 2-way SPES as the public key which is generated using a totally different evolution operator compared to that of the 3-way SPES based protocol, see Appendix~\ref{appd} . We note that 3-way SPES and nonlocal 2-way SPES are fundamentally different. Furthermore, in step 3, the inverse evolution operator $U^{-1}$ follows the step 1 evolution operator $U$ (which is different for both the protocols), thus step 3 is also different for the two protocols, see Appendix~\ref{appd}.

The cryptographic protocols discussed in this letter, utilizing the genuine 3-way and nonlocal 2-way SPES, are compared with other relevant cryptographic proposals~\cite{me-cb2,crypt15,apanda,cryp13}, see details in Appendix~\ref{appf}. These protocols are more efficient than existing proposals~\cite{me-cb2,crypt15,apanda,cryp13} by offering enhanced security in public-key generation, implementing a resource-efficient message decryption method, exhibiting greater resilience against eavesdropping and facilitating easier Eve detection. It is also worth noting that this letter stands out as the sole proposal capable of securely encoding two distinct messages concurrently.



\section{Photon-based Experimental realization of our protocols}
Our cryptographic protocols based on 3-way and nonlocal 2-way SPES can be experimentally realized with a photonic circuit~\cite{Chandra2022,expt1-science,2dqw-expt,BuschPRL} using a single-photon source, wherein the OAM (orbital angular momentum), path and polarization of a single-photon are mapped respectively to the 2D alternate quantum walker's position $x$-DoF, $y$-DoF and coin-DoF. This experimental setup requires passive optical devices such as polarization beam splitters (PBS) and Jones-plates (J-plates)~\cite{jplate}. A PBS reflects horizontal polarization ($\ket{0_c}$) and transmits vertical polarization  ($\ket{1_c}$), thus, this device acts as $\hat{S}_y$ operator of the AQW dynamics. A J-plate modifies the photon's OAM states depending upon its polarization states, which makes this device imitate the AQW operator sequence: coin-shift in $x$ direction-coin, i.e., $J(\alpha,\beta,\gamma)\equiv[\mathbb{1}_{xy}\otimes \hat{M1}]).\hat{S}_x
.[\mathbb{1}_{xy}\otimes \hat{M1}]$. A schematic representation for this photon-based implementation of the cryptography protocols is given in Fig.~\ref{f3}. The output photon from Bob's setup, as in Fig.~\ref{f3}(b), has 3-way  entanglement (SPE) between its path, OAM, and polarization states, and the entanglement quantifier $\pi$-tangle can be calculated via full state tomography or a machine-learning assisted measurement scheme with a series of swap operators
on copies of the quantum state~\cite{negexpt}. This 3-way entangled single-photon is the public key; Alice then encodes two messages (m,n), one in the OAM state and the other in its path state via an appropriate arrangement of PBS and J-plates (i.e., a $\hat{T}$ operation). After receiving the message-encoded photon, Bob will use a sequence of PBS and J-plates akin to the inverse of the AQW evolution operator, as shown in Fig.~\ref{f3}(c). Then Bob finally recovers the encrypted messages by measuring the  OAM and path states of the photon at his detectors. A single photon detector (SPD) measures the photon's path state, and thus, Bob obtains the message $m$. Bob measures the photon's OAM state using an SPD, a single-mode fiber, and a spatial light modulator at his detectors, thereby obtaining the message $n$. A photonic realization can be done with a public key exhibiting nonlocal 2-way SPE; see Appendix~\ref{appg}.
\\

\section{Conclusion}
A resource-saving 2D AQW setup can generate genuine three-way or nonlocal two-way single photon entangled state(SPES) with a single photon that is initially in a separable (product) state. Using this, we develop cryptographic protocols such that the generated 3-way or nonlocal 2-way SPES encode two distinct messages in the two position degrees of freedom $x$ and $y$ of the quantum-walker (single photon). Herein, both the 3-way and nonlocal 2-way SPES work as cryptographic public keys, and then the encryption-decryption of two sets of messages can be securely enabled between the sender-receiver duo. We also outlined the experimental implementation of the cryptographic schemes using a single photon. Besides being the sole protocol for simultaneously encrypting two distinct messages in a resource-saving manner, this protocol exhibits enhanced security compared to existing ones. Our protocol enables multitasking, e.g., it can be used for secure direct communication of two messages (simultaneously) or as a QKD scheme explicitly in the decryption part (Fig.~2(c)) where Alice (sender) transmits a random string of numbers, i.e., two numbers in each shot. A cryptographic protocol exploiting more than three DoF of a single quantum particle, such as a single photon or a trapped ion~\cite{ion1,ion2}, to encrypt and decrypt more than two messages simultaneously can be an extension of this proposal. Our presented work significantly contributes towards the forefront of quantum cryptography and communication by leveraging the novel resource-saving alternate 2D AQW setups. 


\section{Appendix: }
\subsection{Entanglement quantifiers for 2-way and genuine 3-way single-particle entanglement (SPE)}
\label{appa}

The amount of genuine 3-way SPE in quantum states of a particle evolving via 2D alternate quantum walk (AQW) dynamics initialized with
arbitrary separable states (as discussed in the main text) are the same as the degree of quantum correlations between the particle's three degrees of freedom ($x,y$ and coin). This 3-way entanglement can be measured by a negativity-based monotone, $\pi$-tangle~\cite{fan,pitang}, denoted by, $\pi_{xyc}$. It is defined as,
\begin{equation}
\begin{aligned}
   & \pi_{xyc}=\frac{\pi_x+\pi_y+\pi_c}{3}, \\& \text{where, } \pi_x=N_{x|yz}^2- (N_{xy}^2+N_{xz}^2),\;\\&
    \pi_y=N_{y|zx}^2- (N_{yz}^2+N_{yx}^2),\;\\&
    \pi_z=N_{z|xy}^2- (N_{zx}^2+N_{zy}^2).    
\end{aligned}
\label{eq7}
\end{equation}
Here, $N_{i|jk}=\frac{||\rho^{\text{T}_i}_{ijk}||-1}{2}$ and $N_{ij}=\frac{||\rho^{\text{T}_i}_{ij}||-1}{2},$ where $i,j,k \in \{x,y,c\}$.  $\rho_{ijk}=\ket{\psi(t)}\bra{\psi(t)}$, i.e., the density operator for the evolved state $\ket{\psi(t)}$ (i.e., $\ket{\psi_t}$ shown in main text Eq.~(5)) of the quantum walker and $\rho_{ij}=\text{Tr}_k(\rho_{ijk})$ is a reduced density operator after tracing out $i$ degree of freedom (DoF). Also, $\text{T}_i$ stands for the partial transpose concerning $i$-DoF and $||.||$ denotes the matrix trace norm.
Note that $\pi_{xyc}\ge0$ in general, but $\pi_{xyc}=0$ in the absence of genuine 3-way SPE~\cite{fan} in the particle, e.g., for separable or product states such as the initial state shown in main text Eq.~(1). 
\vspace{0.5cm}

Moreover, the amount of nonlocal two-way SPE in a single particle that evolves via the 2D AQW dynamics, i.e., entanglement between $x$ and $y$ position DoF of the walker (particle), can be quantified by negativity $N$~\cite{neg2002}. To evaluate this, at any arbitrary time step $t$ of the AQW, we perform a partial trace of the full density operator, $\rho_{xyc}=\ket{\psi(t)}\bra{\psi(t)}$, concerning the coin degree of freedom, and we get, $\rho_{xy}=\text{Tr}_c(\rho_{xyc})$. Then the eigenvalues $\lambda_i$ where $i$ is from 1 to $(2t+1)^2$, of the partial-transpose of $\rho_{xy}$ with respect to either $x$-DoF or $y$-DoF, are evaluated. From this, one finds entanglement negativity to be,
\begin{equation}
    N=\sum_{i}\frac{|\lambda_i|-\lambda_i}{2}, \text{where $i$ runs from 1 to $(2t+1)^2$}.
    \label{eq8}
\end{equation}

Note $N$ is positive for entangled states, i.e., for nonlocal 2-way SPES, whereas $N=0$ for non-entangled or product states such as initial states shown in main text Eq.~(1).

\subsection{Proof of commutation relation essential to the cryptography protocol}
\label{appb}
Here we prove the commutation relation, $[U,\hat{T}]=0$ between the general evolution operator $U=\hat{S}_y.[\mathbb{1}_{xy}\otimes \hat{C}].\hat{S}_x
.[\mathbb{1}_{xy}\otimes \hat{C}]$ (see main text Eq.~(4))
 and the operator $\hat{T}=\sum_{i,j=-t}^{t}\ket{(i+m)_x}\bra{i_x}\otimes\ket{(j+n)_y}\bra{j_y}\otimes\mathbb{1}_c$ used for the message ($m,n$) encryption (see main text page-3) as follows. Let $U=U_y.U_x$ where $U_y=\hat{S}_y.[\mathbb{1}_{xy}\otimes \hat{C}]$ and $U_x=\hat{S}_x
.[\mathbb{1}_{xy}\otimes \hat{C}]$. 
As shown in main text Eq.~(2), the general coin operator is, 
\begin{equation}
\hat{C}(\alpha,\beta,\gamma) =
\begin{pmatrix}
\cos{(\alpha)} & e^{i\beta} \sin{(\alpha)}\\
\sin{(\alpha)}e^{i\gamma} & -e^{i(\beta+\gamma)} \cos{(\alpha)}
\end{pmatrix},
\label{eq9}
\end{equation}
where $\alpha,\beta,\gamma\in[0,2\pi]$ and the shift operators (conditional), as given in main text Eq.~(4), are,
\begin{equation}
\begin{aligned}
\hat{S_x} = & \sum_{j=-\infty}^{\infty}\{\ket{(j-1)_x}\bra{j_x}\otimes\mathbb{1}_y\otimes\ket{0_c}\bra{0_c}\\
&+\ket{(j+1)_x}\bra{j_x}\otimes\mathbb{1}_y\otimes\ket{1_c}\bra{1_c}\},
\\ \text{and }
\hat{S_y} = & \sum_{j=-\infty}^{\infty}\{\mathbb{1}_x\otimes\ket{(j-1)_y}\bra{j_y}\otimes\ket{0_c}\bra{0_c}\\
&+\mathbb{1}_x\otimes\ket{(j+1)_y}\bra{j_y}\otimes\ket{1_c}\bra{1_c}\}.
\end{aligned}
\label{eq10}
\end{equation} 
Here the coin operator $\hat{C}$ of interest, is $\hat{M1}=\hat{C}(\alpha=\frac{5\pi}{16},\beta=\frac{\pi}{2},\gamma=\frac{\pi}{2})$ for 3-way SPES (single-particle entangled states) generation or $\;\hat{G1}=\hat{C}(\alpha=\frac{19\pi}{16},\beta=\frac{\pi}{2},\gamma=\frac{\pi}{2})$ for nonlocal 2-way SPES generation, see main text page-2 and 3.

Clearly, 
 for an arbitrary state vector $\ket{\psi}=\ket{l_x,k_y,q_c}$ of the quantum particle evolving via 2D AQW, we have,  $U_x\ket{\psi}=\epsilon_{0(q)}\ket{(l-1)_x,k_y,0_c}+\epsilon_{1(q)}\ket{(l+1)_x,k_y,1_c}$, where $\epsilon_{0(q)}$ (or, $\epsilon_{1(q)}$) are the amplitudes for the particle being at position $l-1$ (or, $l+1$) in the $x$-direction, depending on the coin state $\ket{q_c}$. For example if $q=0$, then $\epsilon_{0(0)}=\cos(\alpha)$ and $\epsilon_{1(0)}=e^{i\gamma}\sin(\alpha)$. Similarly, we have $U_y\ket{\psi}=\epsilon^{'}_{0(q)}\ket{l_x,(k-1)_y,0_c}+\epsilon^{'}_{1(q)}\ket{l_x,(k+1)_y,1_c}$, where $\epsilon^{'}_{0(q)}$ (or, $\epsilon^{'}_{1(q)}$) are probability amplitudes to locate the particle at position site $k-1$ (or, $k+1$) in the $y$-direction, depending on the coin state $\ket{q_c}$.
 
For the arbitrary quantum state $\ket{\psi}$, 
 \begin{align}
 \begin{split}
     \hat{T}U\ket{\psi}=&\hat{T}U_y[\epsilon_{0(q)}\ket{(l-1)_x,k_y,0_c}+\epsilon_{1(q)}\ket{(l+1)_x,k_y,1_c}]\\
     &=\hat{T}[\epsilon_{0(q)}(\epsilon^{'}_{0(0)}\ket{(l-1)_x,(k-1)_y,0_c} \\&\;\;\;+\epsilon^{'}_{1(0)}\ket{(l-1)_x,(k+1)_y,1_c})\\&\;\;\;+\epsilon_{1(q)}(\epsilon^{'}_{0(1)}\ket{(l+1)_x,(k-1)_y,0_c} \\&\;\;\;+\epsilon^{'}_{1(1)}\ket{(l+1)_x,(k+1)_y,1_c})]\\&
     =\epsilon_{0(q)}(\epsilon^{'}_{0(0)}\ket{(l-1+m)_x,(k-1+n)_y,0_c} \\&\;\;\;+\epsilon^{'}_{1(0)}\ket{(l-1+m)_x,(k+1+n)_y,1_c})\\&\;\;\;+\epsilon_{1(q)}(\epsilon^{'}_{0(1)}\ket{(l+1+m)_x,(k-1+n)_y,0_c} \\&\;\;\;+\epsilon^{'}_{1(1)}\ket{(l+1+m)_x,(k+1+n)_y,1_c})
 \end{split}
 \label{eq11}
 \end{align}

 On the other hand, we find, 
 \begin{align}
     \begin{split}
         U\hat{T}\ket{\psi}=&U_yU_x[\ket{(l+m)_x,(k+n)_y,q_c}]\\&
         =U_y[\epsilon_{0(q)}\ket{((l+m)-1)_x,(k+n)_y,0_c}\\&\;\;\;+\epsilon_{1(q)}\ket{((l+m)+1)_x,(k+n)_y,1_c}]\\&
        =\epsilon_{0(q)}(\epsilon^{'}_{0(0)}\ket{(l+m-1)_x,(k+n-1)_y,0_c}\\&\;\;\; +\epsilon^{'}_{1(0)}\ket{(l+m-1)_x,(k+n+1)_y,1_c})\\&\;\;\;+\epsilon_{1(q)}(\epsilon^{'}_{0(1)}\ket{(l+m+1)_x,(k+n-1)_y,0_c} \\&\;\;\;+\epsilon^{'}_{1(1)}\ket{(l+m+1)_x,(k+n+1)_y,1_c})
     \end{split}
     \label{eq12}
 \end{align}
 From close observation of Eqs.~(\ref{eq11}) and (\ref{eq12}), we see that $U\hat{T}\ket{\psi}=\hat{T}U\ket{\psi}$ for any arbitrary state $\ket{\psi}$. This proves that $U\hat{T}=\hat{T}U$, i.e., $[U,\hat{T}]=0$.

To justify the essential use of this commutation relation in the quantum cryptography protocol mentioned in the main text, let us look at Step 3 (Message Decryption) by Bob as discussed in main text page-3, where Bob acts the operator $\hat{M}=(U^{-1})^t$ on Alice's message-encoded quantum state $\ket{\psi(m,n)}=\hat{T}\ket{\psi_{pk}}$ where the public key, $\ket{\psi_{pk}}=U^t\ket{l_x,k_y,q_c} $. Bob obtains, 
\begin{equation}
\begin{aligned}
    \hat{M}\ket{\psi(m,n)}= &  (U^{-1})^t\hat{T}\ket{\psi_{pk}} =(U^{-1})^t\hat{T}U^t\ket{l_x,k_y,q_c} \\ & = U^{-t}\hat{T}U^1U^{t-1}\ket{l_x,k_y,q_c}\\ &= U^{-t}U^1\hat{T}U^{t-1}\ket{l_x,k_y,q_c}\\ & =...=U^{-t}U^{t-1}\hat{T}U^1\ket{l_x,k_y,q_c}\\ & =U^{-t}U^t\hat{T}\ket{l_x,k_y,q_c}\\ & =\ket{(l+m)_x}\ket{(k+n)_y}\ket{q_c}.
\end{aligned}
\label{eq13}
\end{equation}
From this outcome of Eq.~(\ref{eq13}), Bob securely reads the two messages ($m,n$) encoded by Alice. 
The second and third lines of Eq.~(\ref{eq13}) justify the necessity of the commutation relation $[U,\hat{T}]=0$ in the cryptographic protocol. In other words, without the commutation relation, message decryption is impossible.

\subsection{Proof for security of the public key ($\ket{\psi_{pk}}$) and the message-containing quantum state ($\ket{\psi(m,n)}$)}
\label{appc}
Herein, we prove the unconditional security of the public key $\ket{\psi_{pk}}$ (and hence the message-containing quantum state $\ket{\psi(m,n)}$) using Holevo's Theorem, which limits the amount of classical information an eavesdropper (Eve) can extract from a quantum mixed state via using a POVM measurement~\cite{crypt15,hol1,hol2,hol3}. As Eve a priory does not know the private key (i.e. $\{U, l_x,k_y,t\}$), the public key($\psi_{pk}$) perceived by Eve is a mixed state $\rho_{pk}$. Even if Eve were aware of $U^t$ (which itself practically has nearly negligible probability seeing the infinite possible combinations for the complex 2D AQW evolution operator $U$), $\rho_{pk}$ still remains entirely mixed, i.e., 
\begin{equation}
\begin{aligned}
\rho_{pk} 
	=& U^t\{\frac{1}{2(2N+1)^2} \sum_{l = -N}^{N} \sum_{k =  -N}^{N}\sum_{q\in\{ 0,1\}} \Ket{l_x}  \Bra{l_x}  \\& ~~~~~~~~\otimes \Ket{k_y} \Bra{k_y} \otimes \Ket{q_c} \Bra{q_c} \}(U^t)^{\dagger}\\
	=& U^t \left(\frac{1}{2(2N+1)^2} \mathbb{1}_{x} \otimes \mathbb{1}_{y} \otimes \mathbb{1}_c\right) (U^t)^{\dagger}
	\\
	=& \frac{1}{2(2N+1)^2}\; \left( \mathbb{1}_{x} \otimes \mathbb{1}_{y}\otimes \mathbb{1}_c \right).
\end{aligned}
\label{sec1}
\end{equation}
Note that here, we considered the coin ket $\ket{q_c}$ with $q\in\{ 0,1\}$ for simplicity of the security proof, but indeed, it can be any arbitrary superposition as shown in our protocol; see page 3 above.
Let Eve perform a measurement on $\rho_{pk} $ to extract any information, Holevo's theorem dictates that the mutual information $I(pvk,eo)$ between the private key ($pvk$) and her measurement outcome or deduction ($eo$) is constrained by the von Neumann entropy of the state, i.e.,
\begin{equation}
\begin{aligned}
I(pvk,eo) \leq & S_{von}(\rho_{pk})=-\Tr(\rho_{pk} \log_2\rho_{pk}),\\ &\leq \log_2(2(2N+1)^2))\; 
\\&\leq 1+2\log_2(2N+1)\;
\end{aligned}
\label{sec2}
\end{equation}
The cryptographic protocol or the public key is secure, if the Shannon entropy of the private key $H(P_{pvk})$ is very large compared to the mutual information $I(pvk,eo)$.
The Shannon entropy of the private key is dependent on the probability of selecting each element from the private-key set $\{U, l_x, k_y, t\}$. Here, we denote by $p_U$ the probability of choosing the AQW evolution operator $U$ from the set $U_{set}=\{U_i=U(\alpha_i,\beta_i,\gamma_i)~|~i\in [1,D]\}$, with considering the natural number $D$ as infinitely large as possible ($D\rightarrow\infty$) since $U$ depends on the choice of the coin operator $\hat{C}(\alpha_i,\beta_i,\gamma_i)$ with $\alpha_i,\beta_i,\gamma_i \in [0,2\pi]$. We then denote by $p_t$ the probability of running the alternate quantum walk for $t$ time-steps, where $t \in \tau = \{t_1, \dots, t_{max}\}$, and by $p_{l,k,q}$ the probability of choosing $\Ket{l_x}\ket{k_y}\ket{q_c}$ as an initial state of the quantum walker (e.g., a single-photon), where $l,k \in \{-N,-(N-1),...,-2,-1,0,1,2,..., N-1, N\}$ and $q \in \{0,1\}$.

We know that these choices are random and completely independent of each other, the probability of obtaining a specific private key $pvk$ can be expressed as:
\begin{equation}
P_{pvk}=p_{U} \; p_{l,k,q}\; p_t=\frac{1}{D\; ~2(2N+1)^2 |\tau|}.
\end{equation}
Here, $|\tau|$ denotes the number of elements in the set $\tau$.
Since the probability distributions above are uniform, we find the Shannon entropy of the private key as,

\begin{equation}
\begin{aligned}
H(P_{pvk})=&
	-\sum_{U\in U_{set}} \sum_{l = -N}^{N} \sum_{k =  -N}^{N}\sum_{q\in\{ 0,1\}}\sum_{t\in \tau}
\{ p_{U} \; p_{l,k,q}\\& \;\;\;\;\; p_t\log_2(p_{U} \; p_{l,k,q}\; p_t)\}\\[2mm]
	=& \log_2(D\; |\tau| \; 2(2N+1)^2) \\[2mm]
	=& \log_2(D\; |\tau|) + 2\log_2(2N+1)+1.
\end{aligned}
\label{sec4}
\end{equation}
Clearly, from a close observation of Eq.~(\ref{sec2}) and Eq.~(\ref{sec4}), we obtain,
\begin{equation}
H(P_{pvk})=S_{von}(\rho_{pk})+\log_2(D\; |\tau|)
\label{sec5}
\end{equation}
Thus, in other words,
\begin{equation}
H(P_{pvk})>>S_{von}(\rho_{pk})\ge
I(pvk,eo),
\end{equation}
which proves our aim that the public key is secure against eavesdropping. We can make an appropriate choice for $D$ and $|\tau|$, such that $\log D\;, |\tau|\approx \text{poly}(\nu)$ for a large $\nu$ value, where $\nu=\text{log}_2(2 N+1)$ refers to the maximum number of bits corresponding to each of two messages, $m,n\in\{-N,-(N-1),...,-2,-1,0,1,2,..., N-1, N\}$.
In this context from Eq.~(\ref{sec5}), the Shannon entropy of the private key exhibits at least a polynomial increase relative to the von Neumann entropy of the public key as perceived by Eve, i.e.,
\begin{equation}
H(P_{pvk})-S_{von}(\rho_{pk})\approx \text{poly}(\nu)\log_2(\text{poly}(\nu)).
\end{equation}
Therefore, even if Eve extracts the maximum possible information about the private key (or the public key), quantified by $S_{von}(\rho_{pk})$, her uncertainty (measured in bits) about the private key $pvk$ remains at least polynomial in $\nu$. Therefore, the public key is unconditionally secure against Eve's attacks. It is straightforward to see that without the knowledge of the public key (or the private key), the message-containing quantum state $\ket{\psi(m,n)}$ as perceived by Eve is also a completely mixed state, and it is at least as secure as the public key (proven above to be secure). Furthermore, as a direct consequence of the fact that a specific private key (which is just proven to be secure) can only enable the decryption of the messages $(m,n)$ from the dual-message encrypted quantum state $\ket{\psi(m,n)}$, Eve figuring out the messages is impossible.

\subsection{Nonlocal 2-way SPES based quantum-cryptographic protocol}
\label{appd}
The nonlocal 2-way SPES generated via the 2D AQW method can also be used to design a quantum cryptographic protocol to encode two messages ($m,n$) in the two nonlocal DoFs ($x,y$) of the quantum particle evolving via the AQW. We discuss the protocol considering the optimal entangler $G1_xG1_y...$ in the subsequent steps--

\underline{Step 1}: Consider, Alice plans to transfer two messages  $m$ and $n$ to Bob securely, where $m,n\in\{-N,-(N-1),...,-2,-1,0,1,2,...,N-1,N\}$. Here, $N$ is the maximum value for the position of the walker in the  $x$ or $y$ direction at time step $t$ in the AQW evolution. Bob generates the public key as $\ket{\phi_{pk}}=U^t\ket{l_x}\ket{k_y}\ket{q_c}$, where $U=\hat{S}_y.[\mathbb{1}_{xy}\otimes \hat{G1}].\hat{S}_x
.[\mathbb{1}_{xy}\otimes \hat{G1}]$. $\ket{l_x}$ and $\ket{k_y}$ are respectively the $x$ and $y$ position states and $\ket{q_c}=\cos(\frac{\theta}{2})\ket{0_c}+e^{i\phi}\sin(\frac{\theta}{2})\ket{1_c}$ be the coin state, of the particle. This public-key is entangled in the nonlocal $x$ and $y$-DoF, for example, with an entanglement value $N_{xy}=0.4273$ at $t=2$, $N_{xy}=2.9398$ at $t=10$ for  $\phi=\pi$ and $\theta=\frac{\pi}{2}$ at both time-steps (see main text Fig.~1(b)). Bob then sends this public key to Alice at any arbitrary AQW time-step ($t$).

\underline{Step 2}: Alice encodes the messages $m$ and $n$ respectively in the $x$-DoF and $y$-DoF, via $\ket{\psi(m,n)}=\hat{T}\ket{\psi_{pk}}$, where $\hat{T}=\sum_{i,j=-N}^{N}\ket{(i+m)_x}\bra{i_x}\otimes\ket{(j+n)_y}\bra{j_y}\otimes\mathbb{1}_c$. Alice then sends this encoded message to Bob via the quantum state $\ket{\psi(m,n)}$.

\underline{Step 3}: Bob decodes Alice's message via the operation $\hat{M}=U^{-t}$ on $\ket{\psi(m,n)}$. It is again essential to have $[U,\hat{T}]=0$, i.e., the operators $U$ and $\hat{T}$ commute, for proceeding further with the cryptographic protocol. The detailed proof of the commutation relation $[U,\hat{T}]=0$ is provided in Appendix~\ref{appb} above. Due to the commutation relation, Bob obtains, 

\begin{equation}
\begin{aligned}
\hat{M}\ket{\psi(m,n)}=& U^{-t}\hat{T}\ket{\psi_{pk}}=U^{-2}\hat{T}U^t\ket{l_x,k_y,q_c}\\&=U^{-t}U^1\hat{T}U^{t-1}\ket{l_x,k_y,q_c}\\&=...=U^{-t}U^t\hat{T}\ket{l_x,k_y,q_c}\\&=\ket{(l+m)_x}\ket{(k+n)_y}\ket{q_c}.
\end{aligned}
\label{eq14}
\end{equation}

Finally, Bob securely reads the two messages ($m,n$) encoded by Alice from this outcome. 

The security of the cryptographic scheme against any eavesdropper attack is discussed below in Appendix~\ref{appe}. 

\subsection{Security of the nonlocal 2-way SPES-based quantum cryptographic protocol with example messages}
\label{appe}

Here, we discuss the cryptography scheme based on 2-way SPES and its resilience against any eavesdropping attacks, with example messages of $m=1$ and $n=2$.

Let $\ket{l_x}=\ket{0_x}\;,\ket{k_y}=\ket{0_y}\;,\ket{q_c}=\frac{1}{\sqrt{2}}(\ket{0_c} -\ket{1_c})$. These correspond to the initial state (separable) as shown in main text Eq.~(1), i.e., $\ket{\psi_0} = \ket{0_x,0_y}\otimes[\cos(\frac{\theta}{2})\ket{0_c} + e^{i\phi}\sin(\frac{\theta}{2})\ket{1_c}]$,  with ($\phi=\pi,\;\theta=\frac{\pi}{2}$), and let us consider the AQW state at time-step $t=2$ for simplicity in the calculations.

\underline{Step 1}: The public key generated by Bob is, $\ket{\phi_{pk}}=U^2\ket{0_x}\ket{0_y}\frac{1}{\sqrt{2}}(\ket{0_c} -\ket{1_c})$, where $U=\hat{S}_y.[\mathbb{1}_{xy}\otimes \hat{G1}].\hat{S}_x
.[\mathbb{1}_{xy}\otimes \hat{G1}]$.

Thus, 
\begin{equation}
\begin{aligned} 
\ket{\phi_{pk}}=&U^2\ket{0_x,0_y}\otimes\frac{1}{\sqrt{2}}(\ket{0_c}-\ket{1_c})\\&= \frac{1}{8192}[v^3u^{-2}(-8e^{\frac{i\pi}{2}}+\sqrt{2}vu^{-2})\ket{-2_x,-2_y,0_c} \\&\;\;\;+e^{\frac{i\pi}{2}}\{32e^{\frac{i\pi}{2}}v(\sqrt{2}v-8e^{\frac{i\pi}{2}}u^{2})\ket{-2_x,0_y,0_c}
\\&\;\;\;+(64\sqrt{2}e^{\frac{i\pi}{2}}v^2+8e^{i\pi}v(-32+v^2)u^{-4})u^2)\ket{0_x,-2_y,0_c}\\&\;\;\; -(64\sqrt{2}e^{\frac{3i\pi}{2}}v^2+8e^{2i\pi}v^3u^{-2}-256e^{2i\pi}vu^{2})\ket{0_x,0_y,0_c} \\&\;\;\;+(1024\sqrt{2}e^{\frac{3i\pi}{2}}u^{4}+256e^{2i\pi}vu^{2})\ket{2_x,-2_y,0_c}\\&\;\;\;+(32\sqrt{2}e^{\frac{5i\pi}{2}}v^2+8e^{3i\pi}v^3u^{-2})\ket{2_x,0_y,0_c}
\\&\;\;\;-(32\sqrt{2}e^{\frac{i\pi}{2}}v^2-8v^3u^{-2})\ket{-2_x,0_y,1_c}
+(256e^{i\pi}vu^{2}\\&\;\;\;-1024\sqrt{2}e^{\frac{3i\pi}{2}}u^{4})\ket{-2_x,2_y,1_c} \\&\;\;\;+(64\sqrt{2}e^{\frac{3i\pi}{2}}v^2-8e^{i\pi}v^3u^{-2}+256e^{i\pi}vu^2)\ket{0_x,0_y,1_c} \\&\;\;\; -(64\sqrt{2}e^{\frac{5i\pi}{2}}v^2-8e^{2i\pi}v^3u^{-2}+256e^{2i\pi}vu^2)\ket{0_x,2_y,1_c} \\&\;\;\;-(32\sqrt{2}e^{\frac{5i\pi}{2}}v^2+256e^{2i\pi}vu^{2})\ket{2_x,0_y,1_c}\\&\;\;\; -(\sqrt{2}e^{\frac{7i\pi}{2}}v^4u^{-4}+8e^{3i\pi}v^3u^{-2})\ket{2_x,2_y,1_c}\}] \;\;,
\label{eq15}
\end{aligned}
\end{equation} 
where $u=\sin(\frac{3\pi}{16})$ and $ v=\csc(\frac{\pi}{8})$. This public-key quantum state $\ket{\phi_{pk}}$ with the intertwined complex amplitudes is entangled in the nonlocal $x$ and $y$-DoF, with degree of entanglement $N=0.4273$ (negativity). Bob then sends this public key to Alice. An eavesdropper (Eve) can interrupt the communication in step 2 when Alice transfers the encrypted quantum state $\ket{\psi(m=1,n=2)}$ to Bob.

\underline{Step 2}: Alice encodes the messages $m=1$ and $n=2$ respectively in the nonlocal $x$-DoF and $y$-DoF, via $\ket{\psi(m=1,n=2)}=\hat{T}\ket{\psi_{pk}}$, where $\hat{T}=\sum_{i,j=-2}^{2}\ket{(i+1)_x}\bra{i_x}\otimes\ket{(j+2)_y}\bra{j_y}\otimes\mathbb{1}_c$. Then Alice sends this message containing the quantum state to Bob.

\underline{Step 3}:  In the absence of Eve, Bob receives $\ket{\psi(m=1,n=2)}$ untampered and decodes the Alice messages by operating $\hat{M}=U^{-2}$ on $\ket{\psi(m=1,n=2)}$. As a consequence of the commutation relation $[U,\hat{T}]=0$, (see Appendix~\ref{appb} above), Bob obtains, 
\begin{equation}
\begin{aligned}        
\hat{M}\ket{\psi(1,2)}&=U^{-2}\hat{T}\ket{\psi_{pk}}\\&=U^{-2}\hat{T}U^2\ket{0_x,0_y}\otimes\frac{1}{\sqrt{2}}(\ket{0_c} -\ket{1_c})\\&=U^{-2}U^2\hat{T}\ket{0_x,0_y}\otimes\frac{1}{\sqrt{2}}(\ket{0_c} -\ket{1_c})\\&=\hat{T}\ket{0_x,0_y}\otimes\frac{1}{\sqrt{2}}(\ket{0_c} -\ket{1_c})\\&=\ket{(0+1)_x,(0+2)_y}\otimes\frac{1}{\sqrt{2}}(\ket{0_c} -\ket{1_c})\\&=\ket{1_x}\ket{2_y}\otimes\frac{1}{\sqrt{2}}(\ket{0_c} -\ket{1_c}). 
\label{eq16}
\end{aligned}
\end{equation} Thus Bob got the product state $\ket{1_x}\otimes\ket{2_y}\otimes\frac{1}{\sqrt{2}}(\ket{0_c} -\ket{1_c})$ and finally, from this output state (see, its position kets), Bob securely reads the two messages ($m=1,n=2$) encoded by Alice.

Now, let us consider that Eve is present at Step 2 and has access to quantum computers (say, Q-Eve) or classical computers (say, C-Eve). However, Eve does not know the private key, i.e., the values of $\{U, l_x,k_y,t\}$, thus C-Eve learning the state $\ket{\psi(m=1,n=2)}$ is well-nigh impossible~\cite{crypt15}, also see Appendix~\ref{appc} for its detailed proof. It mitigates the intercept-and-resend Eve attack~\cite{qkd-attack2020} where Eve tries to measure the message-encrypted state or signal, i.e., $\ket{\psi(m,n)}$. Similarly, Q-Eve can not affect, and her measurements perturb and alter the quantum state $\ket{\psi(m,n)}$ and make the state unusable. This attempt of interception will be recognized immediately by the sender-receiver duo. 
Further, the genuine nonlocal 2-way entanglement in public key ($\ket{\psi_{pk}}$) (as compared to a product-state or superposed public key of Refs.~\cite{apanda,crypt15}) makes it even harder for both Q-Eve or C-Eve to extract any information from $\ket{\psi_{pk}}$ (the public key) via POVM measurements~\cite{crypt15}. Additionally, the fact that the private key $\{U, l_x,k_y,t\}$ is shared among Alice and Bob takes care of the risk of the receiver's authentication (i.e., a man-in-the-middle attack~\cite{qkd-attack2020}), i.e., whether the receiver is the friend (Bob) or a foe like Eve. Finally, Q- or C-Eve needs to have information on the operator to be applied on the state $\ket{\psi(m=1,n=2)}$ to retrieve the message(s) encrypted by Alice. Additionally, the probability of any Eve guessing correct $\hat{M}$ and retracing back to the quantum state $\ket{\psi(m=1,n=2)}$ is practically negligible because there exist infinite ways to form coin operators, and their combinations for the complex 2D AQW evolution. 

To summarize, the private key information is inaccessible to Q- or C-Eve and Eve has almost zero probability of knowing Alice's encrypted quantum state $\ket{\psi(m,n)}$. In addition, the nonlocal 2-way entanglement in the public key ($\ket{\psi_{pk}}$) adds another layer of protection (because of this, information extraction from the public key via any measurement is impossible). Due to the infinitely large number of possible  $2\times2$ coin operators and their combinations to form a 2D-AQW-evolution operator, Eve figuring out the exact message-decryption operator $\hat{M}$ (see Step 3 above) and retrieving back to the quantum state $\ket{\psi(m,n)}$, is practically impossible. This quantum cryptographic protocol is, therefore, foolproof and resilient against Eve, no matter whether Eve has access to quantum or classical computers. Although this quantum cryptography scheme is secure, implementing methods like hardware adaptation to isolate quantum systems and employing privacy-amplification is highly recommended~\cite{qkd-attack2020}. These measures can further mitigate deviations from the theoretical prediction in the practical implementation of the quantum cryptographic scheme.

\subsection{Comparison between QW-based quantum cryptographic schemes}
\label{appf}
A comparison of this cryptographic scheme (this paper) based on the genuine three-way SPES and the nonlocal two-way SPES via the 2D AQW setup with other existing relevant cryptographic proposals~\cite{me-cb2,crypt15,apanda,cryp13} is shown in Table~\ref{tab}. The proposal of Ref.~\cite{cryp13} uses the DTQW evolution of two particles, whereas the other proposals (including this paper) involve the DTQW evolution of single particles. The existing cryptographic protocols~\cite{me-cb2,crypt15,apanda,cryp13} employ 1D DTQWs to generate their cryptographic keys. Ref.~\cite{cryp13} showed that a cyclic 1D-DTQW could generate random key sequences due to its inherent nonlinear chaotic dynamic behavior but focuses only on image encryption. The protocol of Ref.~\cite{crypt15} uses quantum states generated through cyclic 1D DTQW (CQW) dynamics to generate public keys, whereas Ref.~\cite{apanda} uses periodic quantum states exploiting chaotic vs. ordered 1D CQW dynamics to achieve the same goal. Ref.~\cite{me-cb2} uses recurrent maximally entangled states generated via a 1D CQW in forming public keys.
On the other hand, our proposal uses genuine 3-way (as well as nonlocal 2-way) SPES states generated via 2D AQW dynamics. One spectacular advantage of our proposal is that one can securely encrypt and decrypt two distinct messages simultaneously, unlike the previous protocols~\cite{me-cb2,crypt15,apanda}, where just one message can be communicated between Alice and Bob. It is possible due to our use of the 2D AQW evolution, which exploits the particle's two infinite spatial Hilbert spaces. In contrast, as in the previous protocols, 1D cyclic DTQWs have only one finite spatial Hilbert space. Regarding message decryption, our proposal with small time steps is as resource-saving as the proposals of Refs.~\cite{me-cb2,crypt15,apanda}. Another distinctive advantage of our protocols is their superior security against eavesdropper attacks over other existing proposals. It is leveraged from using the genuine three-way (or nonlocal 2-way) single particle entangled states as public keys and the complexity of the used 2D AQW evolution. The use of 2D AQW dynamics in this letter expands the range and complexity of possible coin or evolution operators beyond what is achievable with 1D DTQW dynamics of existing proposals. It also makes our protocols more secure against quantum and classical computer-based brute-force attacks than the existing proposals, as a more extensive and intricate set of coins or evolution operators correlates with greater security in a cryptographic algorithm. The protocol in Ref.~\cite{me-cb2} is secure due to its entangled public key. However, it can only encode a single message and includes one position Hibert space (i.e., 1D DTQW evolution)—the protocols of Refs.~\cite{crypt15,apanda} lack entanglement in their public keys and may not be as secure as our proposed protocols and that of Ref.~\cite{me-cb2}.

\begin{widetext}
\begin{table*}[h]
\centering
\caption{Comparison between different DTQW-based quantum cryptographic schemes}
\label{tab7}
\resizebox{\textwidth}{!}{%
\begin{tabular}{|c|l|l|l|l|l|}
\hline
\textbf{Properties$\downarrow$/Proposals$\rightarrow$} &
\multicolumn{1}{c|}{\textbf{\begin{tabular}[c]{@{}c@{}} 3-way (nonlocal 2-way) \\entangled states via 2D\\ AQW based protocol\\ (This paper) \end{tabular}}} &
\multicolumn{1}{c|}{\textbf{\begin{tabular}[c]{@{}c@{}} Recurrent entangled states\\via 1D cyclic QW\\ based protocol\\(D. Panda \& C. Benjamin~\cite{me-cb2})\end{tabular}}} &
\multicolumn{1}{c|}{\textbf{\begin{tabular}[c]{@{}c@{}}Periodic states via\\ 1D cyclic QW \\based protocol \\(A. Panda \& C. Benjamin~\cite{apanda})\end{tabular}}} &
\multicolumn{1}{c|}{\textbf{\begin{tabular}[c]{@{}c@{}}\\Non periodic 1D\\ cyclic QW\\ based protocol\\(Vlachou et al.~\cite{crypt15})\end{tabular}}}&
\multicolumn{1}{c|}{\textbf{\begin{tabular}[c]{@{}c@{}}\\Chaotic 1D \\cyclic QW\\ based protocol \\(Yang et al.~\cite{cryp13})\end{tabular}}} \\ \hline
\hline
\textbf{\begin{tabular}[c]{@{}c@{}}Cryptographic (public) key\\ generating method
\end{tabular}} &
\begin{tabular}[c]{@{}l@{}} 2D AQW, genuine three-way \\or, nonlocal two-way entangled \\states generated via the 2D AQW.
\end{tabular}&
\begin{tabular}[c]{@{}l@{}}Recurrent entangled states via\\ 1D CQW.\end{tabular} &
\begin{tabular}[c]{@{}l@{}}Periodic states via 1D \\CQW.\end{tabular} &
\begin{tabular}[c]{@{}l@{}}States via 1D CQW.\end{tabular} &
\begin{tabular}[c]{@{}l@{}}1D CQW of\\ two particles.\end{tabular} \\
\hline
\textbf{\begin{tabular}[c]{@{}c@{}}No. of simultaneous\\ message(s) for\\ encryption-decryption
\end{tabular}} &
\begin{tabular}[c]{@{}l@{}}Two; one with the $x$-DoF\\ and the other with $y$-DoF.
\end{tabular}&
\begin{tabular}[c]{@{}l@{}}One, with $x$-DoF. \end{tabular} &
\begin{tabular}[c]{@{}l@{}}One, with $x$-DoF.\end{tabular} &
\begin{tabular}[c]{@{}l@{}}One, with $x$-DoF.\end{tabular} &
\begin{tabular}[c]{@{}l@{}}Not applicable.\end{tabular} 
\\ \hline

\textbf{\begin{tabular}[c]{@{}c@{}}Message decryption method
\end{tabular}} &
\begin{tabular}[c]{@{}l@{}}Resource saving for working with\\ few time steps in the AQW\\ evolution.\end{tabular} &
\begin{tabular}[c]{@{}l@{}}Resource saving as it\\ exploits periodicity of DTQW\\ and entangled states.\end{tabular} &
\begin{tabular}[c]{@{}l@{}}Resource saving as it\\ exploits periodicity of DTQW.\end{tabular} &
\begin{tabular}[c]{@{}l@{}}Resource consuming as it\\ exploits inverse of the\\ DTQW evolution operator.\end{tabular} &
\begin{tabular}[c]{@{}l@{}}Not applicable.\end{tabular} \\ \hline

\textbf{\begin{tabular}[c]{@{}c@{}}Resilience against\\ eavesdropper (Eve) attack
\end{tabular}} &
\begin{tabular}[c]{@{}l@{}}Highly secure due to genuine 3-way\\ entanglement in public key, secrecy\\ in public key and complexity of\\ 2D AQW evolution operator.\\ Here, it is easy to detect Eve. \end{tabular} &
\begin{tabular}[c]{@{}l@{}}Secure due to entanglement\\ in public key, secrecy in public\\ key and periodicity of 1D DTQW.\\ Here, it is also easy to detect Eve. \end{tabular} &
\begin{tabular}[c]{@{}l@{}} It is more secure than Vlachou et \\al.~\cite{crypt15} due  to the secrecy of\\ the Public key and\\ 1D DTQW periodicity. \end{tabular} &
\begin{tabular}[c]{@{}l@{}}Secure only due to the\\ secrecy of the Public key\\ via 1D DTQW.\end{tabular} &
\begin{tabular}[c]{@{}l@{}}Not applicable.\end{tabular} \\ \hline
\end{tabular}%
}
\label{tab}
\end{table*}
\end{widetext}

\subsection{Experimental realization of cryptographic protocol based on nonlocal 2-way SPES}
\label{appg}
Our quantum cryptographic protocol using the nonlocal 2-way SPES can also be experimentally realized with a photonic circuit~\cite{Chandra2022,expt1-science,2dqw-expt,BuschPRL} using a single-photon source, similar to the photonic realization of 3-way SPES based cryptographic protocol as discussed in the main text. Here again, the OAM (orbital-angular-momentum), path, and polarization of a single photon are mapped correspondingly into the $x$-position, $y$-position, and coin DoF of the quantum walker. The same experimental circuit shown in the main text, Fig.~2, can also implement this 2-way SPES-based protocol. The experimental setup requires passive optical devices such as polarization beam splitters (PBS) and J-plates~\cite{jplate}. A PBS acts as $\hat{S}_y$ operator of the AQW dynamics, as this device reflects horizontal polarization ($\ket{0_c}$) and transmits vertical polarization ($\ket{1_c}$). Since a J-plate modifies a photon's OAM states depending upon its polarization states, this device imitates the AQW operator sequence: coin-shift in $x$ direction-coin, i.e., $J(\alpha,\beta,\gamma)\equiv[\mathbb{1}_{xy}\otimes \hat{G1}]).\hat{S}_x
.[\mathbb{1}_{xy}\otimes \hat{G1}]$. The output photon from Bob's setup, as in main text Fig.~2(b), has nonlocal 2-way  entanglement (SPE) between its path and OAM states, and the entanglement quantifier entanglement-negativity ($N$) can be calculated via a full state tomography or a machine-learning assisted measurement scheme with a series of swap operators on copies of the quantum state~\cite{negexpt}. This nonlocal 2-way entangled single-photon is the public key. When untampered, Alice can use it to encode two messages (m,n), one in the OAM state and the other in its path state, via an appropriate arrangement of J-plates and PBS. After receiving the message-encoded photon, Bob will use a sequence of PBS and J-plates akin to the inverse of the AQW evolution operator, as shown in the main text, Fig.~2(c). Then Bob finally recovers the encrypted messages by measuring the  OAM and path states of the photon using the detectors at his disposal. A single photon detector (SPD) measures the photon's path state, and then Bob obtains the message $m$. Bob measures the photon's OAM state using an SPD, a single-mode fiber, and a spatial light modulator, and he finally obtains the message $n$.

\twocolumngrid


\begin{thebibliography}{1}


\bibitem{cryp5}  S. Wiesner, SIGACT News 15(1) 78 (1983).


\bibitem{cryp7} D. Mayers, J. ACM 48(3) 351 (2001).

\bibitem{cryp8} T. Schmitt-Manderbach, H. Weier, M. Fürst, R. Ursin, F. Tiefenbacher, T. Scheidl, J.
Perdigues, Z. Sodnik, C. Kurtsiefer, J. Rarity, A. N. Zeilinger and H. Weinfurter, Phys.
Rev. Lett. 98 010504 (2007).
\bibitem{bb84}C. H. Bennett and G. Brassard, Quantum cryptography: Public key distribution and coin tossing, Proceedings of IEEE International Conference on Computers, Systems and Signal Processing, volume 175, page 8. New York (1984).

\bibitem{e91}Artur K. Ekert, Quantum cryptography based on Bell’s theorem, Phys. Rev. Lett. 67, 661 (1991). 
\bibitem{crypt15}C. Vlachou et al., Quantum walk public-key cryptographic system, Int. J. Quantum Inf. 13, 1550050 (2015).

\bibitem{2way} Marco Lucamarini and Stefano Mancini, Secure Deterministic Communication without Entanglement, Phys. Rev. Lett. 94, 140501 (2005).
\bibitem{qw93}Y. Aharonov, L. Davidovich, and N. Zagury, Quantum random
walks, Phys. Rev. A 48, 1687 (1993).
\bibitem{qwqudit19}
Taira Giordani, Emanuele Polino, Sabrina Emiliani, Alessia Suprano, Luca Innocenti, Helena Majury, Lorenzo Marrucci, Mauro Paternostro, Alessandro Ferraro, Nicolò Spagnolo, and Fabio Sciarrino, Experimental Engineering of Arbitrary Qudit States with Discrete-Time Quantum Walks, Phys. Rev. Lett. 122, 020503 (2019).
\bibitem{qwunivqi23}
Kazuma Yonezu, Yutaro Enomoto, Takato Yoshida, and Shuntaro Takeda, Time-Domain Universal Linear-Optical Operations for Universal Quantum Information Processing,
Phys. Rev. Lett. 131, 040601 (2023).
\bibitem{qwgenqmeasure23}Generalized Quantum Measurements on a Higher-Dimensional System via Quantum Walks
Xiaowei Wang, Xiang Zhan, Yulin Li, Lei Xiao, Gaoyan Zhu, Dengke Qu, Quan Lin, Yue Yu, and Peng Xue
Phys. Rev. Lett. 131, 150803 (2023).
\bibitem{cryp13} Y. Yang, Q. Pan, S. Sun and P. Xu, Novel Image Encryption based on Quantum Walks, Sci. Rep. 5 7784 (2015).

\bibitem{me-cb2}Dinesh Kumar Panda and Colin Benjamin, Recurrent generation of maximally entangled single-particle states via quantum walks on cyclic graphs, Phys. Rev. A {108}, L020401 (2023).
\bibitem{apanda} A. Panda and C. Benjamin, Order from chaos in quantum walks
on cyclic graphs, Phys. Rev. A 104, 012204 (2021).

\bibitem{p3}Dinesh Kumar Panda and Colin Benjamin, Designing three-way entangled and nonlocal two-way entangled single particle states via alternate quantum walks, arXiv:2402.05080 [quant-ph] (2024).
\bibitem{BuschPRL}C. Di Franco, M. Mc Gettrick, and Th. Busch, Mimicking the Probability Distribution of a Two-Dimensional Grover Walk with a Single-Qubit Coin, Phys. Rev. Lett. 106, 080502 (2011).

\bibitem{Chandra2022} P. A. Ameen Yasir and C. M. Chandrashekar, Generation of hyperentangled states and two-dimensional quantum walks using 
J
 or 
q
 plates and polarization beam splitters,
Phys. Rev. A 105, 012417 (2022).

\bibitem{pitang}Ariadna J. Torres-Arenas, Qian Dong, Guo-Hua Sun, Wen-Chao Qiang and Shi-Hai Dong,
Entanglement measures of W-state in noninertial frames,
Physics Letters B 789, 93-105 (2019).
\bibitem{fan} Yong-Cheng Ou and Heng Fan, Monogamy inequality in terms of negativity for three-qubit states, Phys. Rev. A 75, 062308 (2007).
\bibitem{neg2002}G. Vidal and R. F. Werner, Computable measure of entanglement,
Phys. Rev. A 65, 032314 (2002).

\bibitem{qkd-attack2020}V. Gaur et al., Quantum Key Distribution: Attacks and Solutions,
Proceedings of the International Conference on Innovative Computing \&
Communications (ICICC) (2020).

\bibitem{expt1-science} M. Karski et al., Quantum Walk in Position Space with Single Optically Trapped Atoms, Science 325, 174 (2009).

\bibitem{2dqw-expt}Hao Tang et al. ,Experimental two-dimensional quantum walk on a photonic chip, SCIENCE ADVANCES 4, eaat3174 (2018).
\bibitem{jplate}Robert C. Devlin, Antonio Ambrosio, Noah A. Rubin, J. P. Balthasar Mueller, Federico Capasso, Arbitrary spin-to–orbital angular momentum conversion of light, Science 358, 896-901 (2017).
\bibitem{negexpt} Machine-Learning-Assisted Many-Body Entanglement Measurement
Johnnie Gray, Leonardo Banchi, Abolfazl Bayat, and Sougato Bose
Phys. Rev. Lett. 121, 150503 (2018).
\bibitem{ion1} H. Schmitz, R. Matjeschk, C. Schneider, J. Glueckert, M. Enderlein, T. Huber, T. Schaetz, Quantum walk of a trapped ion in phase space, Phys. Rev. Lett. 103, 090504 (2009).
\bibitem{ion2}F. Zahringer, et al., Realization of a quantum walk with one and two trapped ions, Phys. Rev. Lett. 104, 100503 (2010).
\bibitem{hol2} 
Nielsen M A and Chuang I L, Quantum Computation and Quantum Information (Cambridge University Press) (2000).
\bibitem{hol3} 
A. S. Holevo, Bounds for the quantity of information transmitted by a quantum
communication channel, Probl Peredachi Inf 9 177 (1973).
\bibitem{hol1} 
Diego G Bussandri and Pedro W Lamberti 2020 J. Phys. A: Math. Theor. 53 045302 (2020).




\end{thebibliography}
\end{document}